\documentclass[proceedings]{JHEP} 

\usepackage{epsfig}                   
\newcommand{\Tr}{{\rm Tr}\,}

\newcommand{\CN}{{\cal N}}
  
\newcommand{\CM}{{\cal M}}
\newcommand{\CP}{{\cal P}}
\newcommand{\CF}{{\cal F}}
\newcommand{\CZ}{{\cal Z}} 
\newcommand{\CO}{{\cal O}}

\conference{Nonperturbative Quantum Effects 2000}

\title{Statistical Physics Approach to M-theory Integrals\thanks{Talk presented by W. Krauth }}

\author{Werner Krauth   and Matthias Staudacher \\
        
        \vspace{0.3 cm} 
        
        CNRS-Laboratoire de Physique Statistique, 
        Ecole Normale Sup\'{e}rieure\\
        24, rue Lhomond\\
        F-75231 Paris Cedex 05\\
        E-mail: \email{krauth@lps.ens.fr}
        
        \vspace{0.3 cm}
        
        Albert-Einstein-Institut, 
        Max-Planck-Institut f\"ur Gravitationsphysik\\ 
        Am M\"uhlenberg 1\\ 
        D-14476 Golm\\
        E-mail: \email{matthias@aei-potsdam.mpg.de}} 

\abstract{
We explain the concepts of computational statistical physics which
have proven very helpful in the study of Yang-Mills integrals, an ubiquitous
new class of matrix models.
Issues treated are: Absolute convergence versus Monte Carlo computability
of near-singular integrals, singularity detection by Markov-chain methods,
applications to asymptotic eigenvalue distributions and to numerical
evaluations of multiple bosonic and supersymmetric integrals. 
In many cases already, it has been possible to resolve controversies
between conflicting analytical results using the methods presented here. 
} 


\begin{document}

\section{Introduction}
\label{Introduction}
Recent work in field theory has revealed the existence of an important
new class of gauge-invariant matrix models. At the difference
of the classic Wigner-type models, interest now focusses on integrals
of D non-linearly {\em coupled} matrices $X_{\mu}, \mu=1,\ldots
D$. The $X_{\mu}$ are constructed from the generators $T^A$ of the
fundamental representation of a given  Lie algebra Lie($G$): $X_{\mu}=
X_{\mu}^A T_A$, with $A = 1,\ldots,{\rm dim}(G)$.  The group $G$
may be $SU(N)$, but the orthogonal, symplectic and exceptional
groups have also come under close scrutiny recently. 

These ordinary, multiple Riemann integrals stem from a dimensional reduction
of D-dimen\-sional Euclidean continuum Yang-Mills theory to zero
dimensions. They have important implications, as the integrals
yield the bulk part of the Witten index of supersymmetric quantum mechanical
gauge theories, 
and appear in multi-instanton calculations of large $N$ susy
Yang-Mills theories. Furthermore, they appear in proposed
formulations of string theory (the IKKT model) and M-theory.
It remains to be elucidated whether they contain further non-perturbative
information on gauge theories via the Eguchi-Kawai mechanism.

For the sake of brevity (cf. \cite{kns} for complete definitions), 
we  write down (even for supersymmetric Yang-Mills
theories) only the effective bo\-sonic integral, which is 
obtained after integrating out the $\CN$(= number of supersymmetries) 
Grassmann-valued fermionic  
matrices
\begin{eqnarray}
\lefteqn{\CZ^{\CN}_{D,G} =  \int \prod_{A,\mu} 
\frac{d X_{\mu}^{A}}{\sqrt{2 \pi}} \times  }
\nonumber \\  
& &
e^{ \frac{1}{4 g^2} \Tr
[X_\mu,X_\nu] [X_\mu,X_\nu]}
(\CP\left\{  X_{\mu}^{A} \right\}).
\label{int}
\end{eqnarray}
In this equation, $\CP(\{X\})$ is the Pfaffian of a certain matrix $\CM$, 
which can be constructed from the adjoint representation of the $X_{\mu}$. 

During the last few years, intense effort has been brought to bear
on these integrals, ranging from the rigorous exact solution for
$SU(2)$ \cite{sestern},\cite{smilga} 
to ultra-sophisticated analytical calculations
\cite{mns} which lent support to earlier conjectures \cite{greengut}.

We have initiated a project with the aim to obtain direct
non-perturbative information on these integrals by  numerical Monte
Carlo calculation.  In several cases already, this approach has 
allowed to clarify analytic properties of the integrals, both in
the supersymmetric and the purely bosonic case (where the Pfaffian
in eq.(\ref{int}) is simply omitted).  We have also obtained very
precise values (statistical estimates) of  $\CZ$ for several
low-ranked groups, estimates which were sufficiently precise to
decide between differing analytical conjectures.  The basic strength
of the numerical approach is however to allow the computation of
a wide range of observables (Wilson loops, eigenvalue distributions),
and much work remains to be done.

The integrals in eq.(\ref{int}) resemble partition functions in
statistical physics. Our initial hope was to reduce eq.(\ref{int})
to a standard form, most simply by the transformation
\begin{equation}
\CZ^{\CN}_{D,G} = \int \prod_{\mu,A}
\frac{d X_{\mu}^{A}}{\sqrt{2 \pi}}
e^{- \frac{ X_{\mu}^{A^2}}{2\sigma^2}}
\left[ 
\frac{
\CF(  \left\{   X_{\mu}^{A} \right\}  )
}  
{  
e^{- \sum \frac{ X_{\mu}^{A^2}}{2\sigma^2 }}
}  
\right] 
\label{gaussian},
\end{equation}
where $\CF(  \left\{   X_{\mu}^{A} \right\}  )$ is the integrand
in the second line of eq.(\ref{int}).
In this form, the integral can (in principle) be computed directly using Monte
Carlo methods.  To do so, it would suffice to generate Gaussian
distributed random numbers $X_{\mu}^{A}$, and to average the term
$[\;\;]$ in eq.(\ref{gaussian}) over this distribution.  The
straightforward approach is thwarted by the  fact that the integrals
eq.(\ref{gaussian}) - and, equivalently, eq.(\ref{int}) - only
barely converge, if at all.

The reason for this bad convergence lies in the existence of
``valleys" in the action  in eq.(\ref{int}).  For example, any
configuration of mutually commuting $X_{\mu}$ ($[X_{\mu},X_{\nu}]=0
\; \forall \mu, \nu$)  gives rise to a subspace of matrices with
vanishing action, which stretches out to infinity, and leads to a
large contribution to $\CZ$. There is presently no mathematical
proof that these singularities are integrable (cf. \cite{nishi}
\cite{ikkt2} for perturbative results).  

Very importantly, integrals may exist, without being computable by
straightforward Monte Carlo methods.  This distinction between existence
and (Monte Carlo) computability is so crucial for Yang-Mills
integrals that we present them in the next section in the simplified
context of a  $1-$dimensional integral.

\section{Existence \& Computability }
\label{Existence }
Consider the integral
\begin{equation}
I(\alpha ) = \int_0^1 dx \mu(x) x^{-\alpha} 
\label{1-d-ex} 
\end{equation}
with a constant weight function $\mu(x)=1$, which we introduce for
later convenience.  In this toy problem, the singularity at
$x=0$ plays the role of a valley, as discussed before, in the more
complex Yang-Mills integral.

\FIGURE
{\epsfig{file=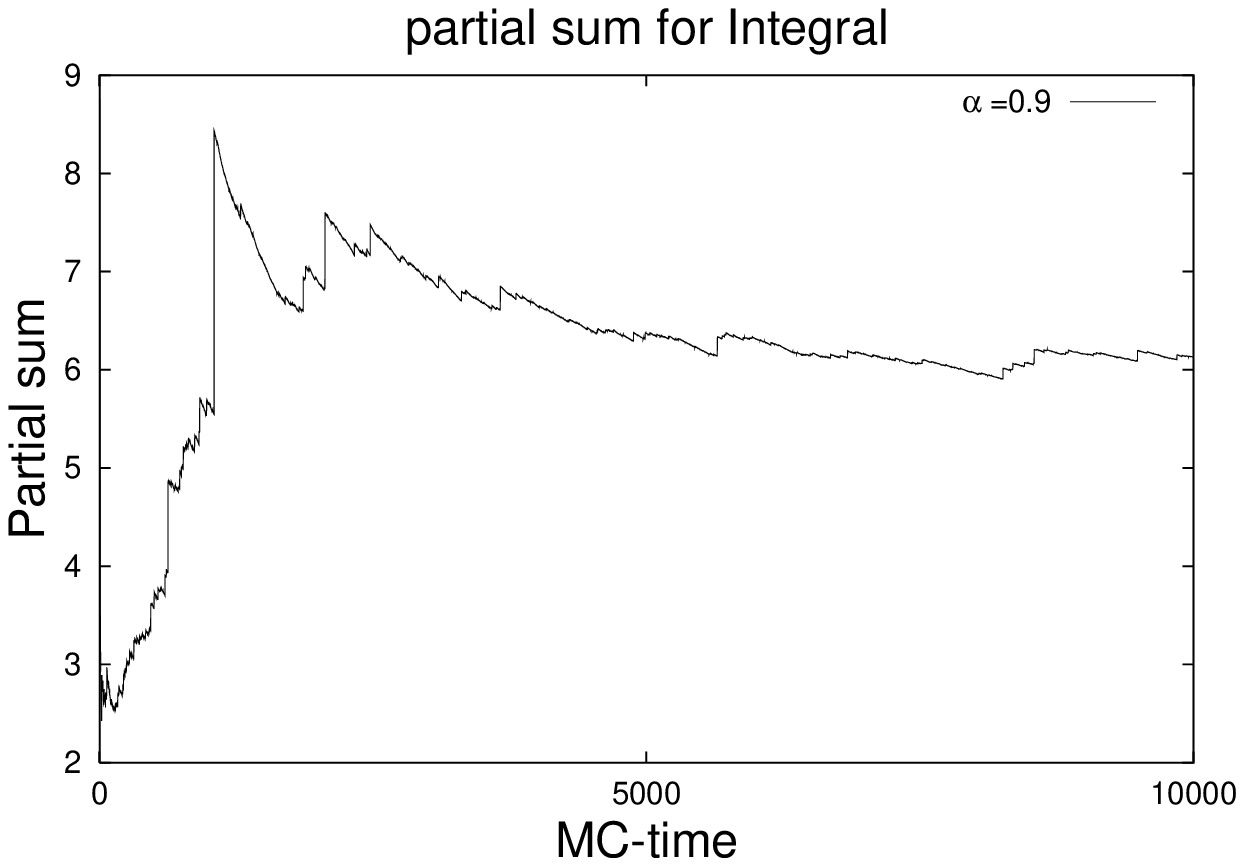,width=6cm}\caption{Partial sums
$S_t$ (cf. eq.(\ref{partial})) for $\alpha=0.9$ vs Monte Carlo time
$t$.  The numerical estimate for this integral seems to converge
to the wrong result.} \label{partialsum0.9}}

We may compute the integral eq.(\ref{1-d-ex}) by the Monte Carlo
method in the following way:  as the weight function is constant
($\mu(x)=1$), we pick $t$ uniformly distributed  points $x_t$ with
$0 \le x_t \le 1$ and compute

\begin{equation}
I \sim S_t = 1/t \sum_{i=1}^t x_i^{-\alpha},    
\label{partial} 
\end{equation}
where $t=1,2\ldots $ is the Monte Carlo time.
A typical outcome for the partial sums $S_t$ during a Monte Carlo
calculation for $\alpha =0.9$ is shown in figure \ref{partialsum0.9}.
The calculation is seemingly correct, as standard
error analysis gives a result $I(\alpha =0.9) = 6.13 \pm 0.46$,
without emitting any warnings! Carrying on the simulation for much
longer times, we would every so often generate an extremely small
$x_t$, which in one step would hike up the partial sum, and change
the error estimate. Repeatedly, we would get tricked into accepting
``stabilized values'' of the integral, which would probably still not
correspond to the true value $I(\alpha =0.9)=10$!

Clearly, there is a problem with the computability of the integral,
which can be traced back to its infinite variance.  Calling $\CO
= x^{-\alpha}$, the variance is given by 
\begin{equation} Var= \int
dx \mu(x) \CO(x)^2 - [\int dx \mu(x) \CO(x)]^2.
\end{equation} 
The error in the Monte Carlo evaluation eq.(\ref{partial}) behaves
like $\sqrt{Var/t}$, and, for  $\alpha \ge  0.5$, is infinite.
This situation  is virtually impossible to diagnose from within
the simulation itself.

We have developed a highly efficient tool to numerically check for
(absolute) convergence of integrals. The idea (translated to the
case of the present toy problem) is to perform Markov chain random
walk simulation with a stationary distribution $\mu'(x)=\mu(x)
x^{-\alpha}$ and $\mu''(x)=\mu(x) x^{-2 \alpha}$.  to check for
existence of the integral and finiteness of the variance, respectively.
In an effort to be completely explicit, this means to choose a
small displacement interval $\delta_t$, uniformly distributed
between $+\epsilon$ and $-\epsilon$, and to go from $x_t$ to
$x_{t+1}$ according to the following probability table

\begin{equation}
x_{t+1} =
\left\{
\begin{array}{cc}
x_{t} + \delta_{t}& \mbox{ w/ probability} \\ 
                 & \mbox{min}(1,\mu'(x_{t+1})/\mu(x_t))   \\ 
x_{t}             & \mbox{else}    \\ 
\end{array} \right.
\label{chain}.
\end{equation}
(cf. \cite{Intro}). 
During these simulations (which are neither used nor useful to compute the
integral eq.(\ref{1-d-ex}) itself), we are exclusively interested in 
finding out whether the Markov chain eq.(\ref{chain}) gets stuck. If
so, it has become attracted  by a point $x_0$ with
\begin{equation}
\int_{x_0}^{x_0+\epsilon} dx \mu'(x) = \infty 
\end{equation}
i. e. a non-integrable singularity.  In figures \ref{supertool0.9}
and \ref{supertool1.8} , we show $x_t$ for Markov chains
with stationary distributions $\mu'(x)$ and $\mu''(x)$, respectively.
Figure \ref{supertool1.8}, in particular, implies that the variance
of the integral eq.(\ref{1-d-ex}) is infinite so that the result
of fig. 1 cannot be trusted, while   figure \ref{supertool0.9}
assures us that the integral exists.

\FIGURE
{\epsfig{file=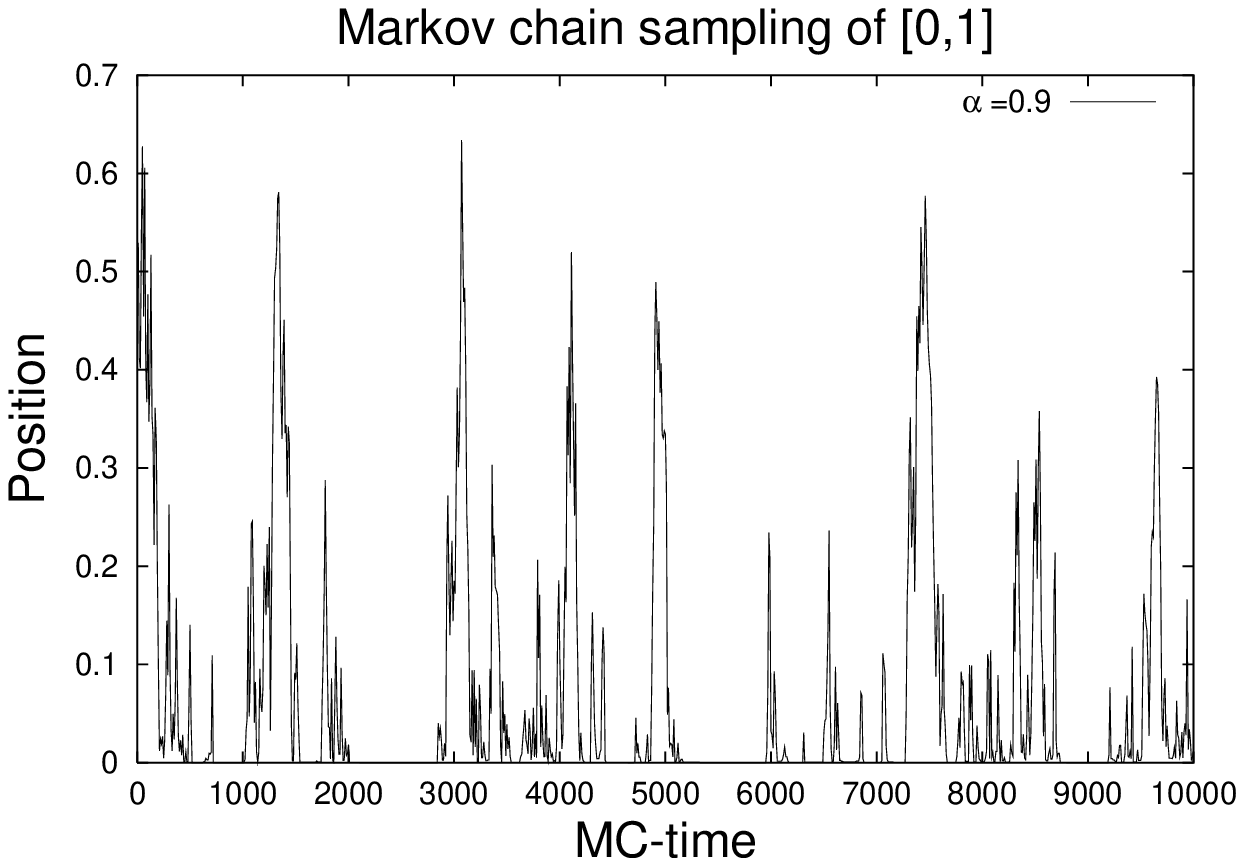,width=6cm}\caption{Position $x_t$ vs
Monte Carlo time $t$ for the Markov chain with stationary distribution
$\mu'(x)= x^{-0.9}$. The time evolution of $x_t$ does not get stuck, 
because the integral $\int_0^1 dx x^{-0.9}$ exists. } 
\label{supertool0.9}}
\FIGURE
{\epsfig{file=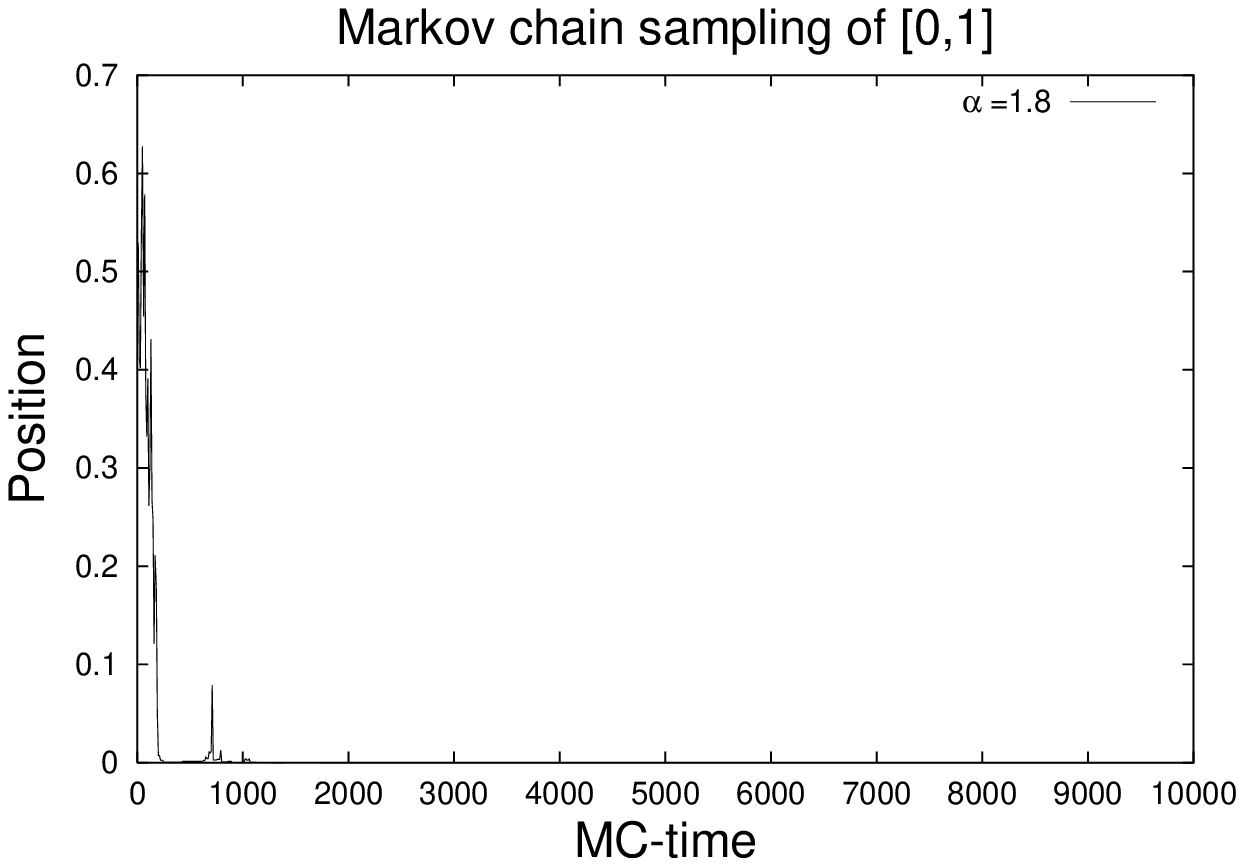,width=6cm}\caption{
Same as figure \ref{supertool0.9}, but for  
$\mu''(x)= x^{-1.8}$. $x_t$ gets stuck at $x \sim 0$ very quickly,
signalling that the variance of the integral eq.(\ref{1-d-ex}) is infinite.
} \label{supertool1.8}}

The method can be  easily adapted to multidimensional integrals
by monitoring an auto-correlation function rather than the position $x_t$,

In our applications, the method has been successful much beyond our initial 
expectations. Besides its ``consulting'' role within the 
Monte Carlo framework (as explained in the caption of figures \ref{supertool0.9}
and \ref{supertool1.8}),
we have used it extensively to establish the existence conditions
for bosonic and susy Yang-Mills integrals, which have not been obtained
analytically beyond the 1-loop level. We have also adopted the method
to obtain important information on  the asymptotic behavior of integrals.

We conclude the discussion of our toy problem by showing how, after all, 
the integral eq.(\ref{1-d-ex}) can be computed by Monte Carlo methods. 
Consider first
\begin{equation}
 Q(\alpha_2,\alpha_1) = \frac{\int_0^1 x^{-\alpha_2 } dx}
{\int_0^1 x^{-\alpha_1 } dx}
  = \frac{\int_0^1
\overbrace{ x^{-\alpha_1 }}^{measure}
\overbrace{ x^{\alpha_1 -\alpha_2 }}^{operator}
 dx}
{\int_0^1 \underbrace{ x^{-\alpha_1 }}_{measure} dx}
\label{meas_op} 
\end{equation}
According to the discussion in section 3,  $Q(\alpha_2,\alpha_1)$ can
be computed from random numbers distributed as $\mu(x)=x^{-\alpha_1} $ 
as long as
\begin{equation}
 2 \alpha_2 - \alpha_1 < 1.
\label{varbound} 
\end{equation}

\TABLE{
\begin{tabular}{||l|l|l|c||} \hline
$\alpha_1$&$\alpha_2$& $ Q(\alpha_2,\alpha_1)$   & error (in \%) \\ \hline
  0.0 &  0.15  &     1.179  &   0.2 \% \\
  0.15 &  0.55 &     1.881  &   1 \%    \\
  0.55 &  0.75 &     1.799  &   1  \%   \\
  0.75 &  0.85 &     1.656  &   0.7 \%  \\
  0.85 &  0.9  &     1.508  &   0.6 \%  \\ \hline
\end{tabular}\caption{Values of $\alpha_1, \alpha_2$, for which the 
quantity $Q(\alpha_2,\alpha_1)$ is computed according to eq.(\ref{meas_op}).
The integral eq.(\ref{1-d-ex}) is well approximated by 
$\prod Q$.} }
It is easy to see that all  of the pairs ($\alpha_1,\alpha_2$) in table 1
satisfy the bound of eq.(\ref{varbound}), and the Monte Carlo data   
for $Q(\alpha_2, \alpha_1)$ can thus be trusted, just as the final result
\begin{equation}
\int_0^1 dx \;\;x^{-0.9} = \prod Q = 9.98 \pm 0.16.
\end{equation}

\section{``Measurement'' = ``Comparison''}
\label{Measurement}

After these preliminary steps, we finally
confront the Monte Carlo measurement of the Yang-Mills integrals.
In this context, we recall from our basic physics training the
heading of this section. 
Translated to the context of a Monte Carlo calculation, the measure/compare
equivalence means that the integral eq.(\ref{int}) has always to be 
written as
\begin{equation}
\CZ^{\CN}_{D,G} =  \int \prod_{A,\mu} 
\frac{d X_{\mu}^{A}}{\sqrt{2 \pi}} \mu(\{X_{\mu}^{A} \}) 
\left\{\frac{\CF(\{X_{\mu}^{A}\}) }{\mu(\{X_{\mu}^{A} \}) } \right\}.   
\label{meas_comp} 
\end{equation}
In the $\{ \;\;\}$ in eq.(\ref{meas_comp}), we {\it compare} $\CF$
to the measure, which we are free to choose (but which we have to
be able to integrate analytically). As mentioned before, the
Gaussians of eq.(\ref{gaussian}) are too different from $\CF$ to
work. A straightforward generalization of the approach eq.(\ref{meas_op}) 
was found to be wanting: the mismatch between $\CF$ and $\mu$ could
only be smoothed with a very large number of steps $(\alpha_1,\alpha_2)$
in eq.(\ref{meas_op}).  

A much better approach has come from the observation that
$\CF$  can be {\it compactified} onto the surface of a hypersphere,
because both the action and the Pfaffian are homogeneous functions
of the radius $R=\sqrt{\sum X_{\mu}^{A}}$, which can therefore be
integrated out.
Introducing polar coordinates $R, \Omega$ and noting 
$\tilde{\CF}=  \int_0^{\infty} dR R^{d-1} \CF(\Omega,R)$, we arrive at 
the ultimate formulation of the integral
\begin{equation}
\CZ^{\CN}_{D,G} = \int d\Omega \tilde{\CF}(\Omega).
\label{hyper} 
\end{equation}
This means that the integrand $ \tilde{\CF}$ is compared to the constant
function on the surface of a hypersphere in dimension $d=dim(G) \CN$ .  

The integral eq.(\ref{hyper}) can still not  be evaluated directly,
so that the strategy of eq.(\ref{meas_op}) has to be used. Here,
we simply compute a few ratios of the integrals $\int d\Omega \left[
\tilde{\CF}(\Omega) \right]^{\alpha}$ for different values $0 <
\alpha < 1$. In this case, of course, pairs $(\alpha_2,\alpha_1)$
are tested by the qualitative Monte Carlo algorithm, as analytical
convergence conditions in the spirit of eq.(\ref{varbound}) are
lacking.  After having expended an extraordinary amount  of rigor
on these very difficult  integrals, we nevertheless obtain
well-controlled predictions, to be surveyed below.

\section{Synopsis}
\label{Synopsis}

The methods presented in the previous sections were used to compute
a number of results which are fully discussed in \cite{kns},
\cite{ks1}, \cite{ks2}, \cite{ks3}. 
For complementary Monte Carlo studies, using somewhat different
techniques, see \cite{nishi}, \cite{aabhn}.
To give an indication of the scope and the quality of the data, we present
here our recent calculations for gauge groups other than $SU(N)$, as well
as an intriguing qualitative result concerning the asymptotics of the 
eigenvalue distributions.

The first example concerns the evaluation of the integrals for
the gauge groups $SO(N), Sp(2N)$ and $G_2$. These calculations can be 
connected to other theoretical work essentially by dividing $\CZ$ 
by the volume $\CF_G$ of the group $G$. In this way, we arrive at a numerical 
value for the bulk contribution to the quantum-mechanical Witten index, 
which is given by 
\begin{equation}
{\rm ind}_0^D(G)=
\frac{1}{\CF_G} \CZ^{\CN}_{D,G}.
\label{bind2}
\end{equation}
 
In table 3 we list our Monte Carlo results for this bulk index,
obtained by the methods explained above, for groups up to rank three.
We furthermore compare these data to analytic predictions from the 
generalization of the deformation method of Moore et al.~\cite{mns}
to these groups \cite{ks3}. Note the excellent precision ($2\%$ 
statistical error) for groups up to $SO(7)$, where the integral 
eq.(\ref{int}) lies in $84$ dimensions. 
Intriguingly, both our numerical and analytical results are
at variance with a previous conjecture \cite{kasm} for non-unitary groups.
In the special case of $SO(7)$, e.g., ref. \cite{kasm} obtains  
the fraction $15/128$ which is incompatible with our data. 
 
\TABLE{
\begin{tabular}
{||c||r@{ $\pm$ }l | r@{   }l  ||} \hline
Group  & \multicolumn{2}{c|}{ Monte Carlo}   & \multicolumn{2}{c||}
{ Exact}      \\
$G$    & \multicolumn{2}{c|}{ ind$^{D=4}_0(G)$}              & \multicolumn{2}
{c||}{ }  \\  \hline \hline SO(3)  & 0.2503  &0.0006
&      & 1/4    \\ SO(4)  & 0.0627  &0.0013
&   & 1/16   \\ SO(5)  & 0.1406  &0.001
&   &  9/64 \\ SO(6)  & 0.0620  &0.001
&   & 1/16   \\ SO(7)  & 0.0966  &0.0017
&   & 25/256 \\ \hline \hline Sp(2)  & 0.2500  &0.0002
&   & 1/4    \\ Sp(4)  & 0.139   &0.0015
&   & 9/64 \\ Sp(6)  & 0.0973  &0.003
&   & 51/512 \\ \hline \hline G$_2$  & 0.173   &0.003
&   & 151/864\\ \hline \end{tabular}
\caption{Monte Carlo results versus proposed (BRST deformation
method) exact values
for the $D=4$ bulk index. } }

Let us mention that at present the calculations for $D=4$ and $D=6$
are considerably simpler than the case $D=10$, because the Pfaffian
can be reduced to a determinant for $D=4,6$ \cite{kns}.  In $D=10$,
this possibility does not exist generically (for an exception for
$SU(3)$ cf. \cite{kns}). We have now developed new methods to
compute Pfaffians which should allow computations 
for $D=10$ in the near future. It is possible if tedious to work out
the predictions of the BRST deformation technique 
for ind$^{D=10}_0(G)$ {\it cf} \cite{indten}, which again differ from
the conjectures of \cite{kasm}. It would be interesting to check
the results of \cite{indten} by our Monte Carlo methods.

A further strength of the Monte Carlo approach is to allow the calculation
of quantities other than just the integral $\CZ$. We briefly review
as a second illustration of the here advocated approach the study of the
correlation functions $< \Tr  X_{\mu}^k\}  >$,
where $X_\mu$ is an arbitrary single matrix. This correlation function
allows to infer the eigenvalue distribution of the matrices. 
Indeed, denoting the normalized eigenvalue density of individual matrices by
$\rho(\lambda)$, one has
\begin{equation}
 < \Tr  X_{\mu}^k\}  > =
\int_{-\infty}^{\infty} d \lambda~\rho(\lambda)~\lambda^k. 
\label{corr} 
\end{equation}
Here 
the calculation was immediately feasible also for $D=10$ case, since
we only needed to test for absolute convergence, i.e.~it suffices to
consider a simplified measure obtained from the absolute value of the 
original measure:
\begin{equation}
\mu'(\left\{X_{\mu}^A\right\}  ) = \Tr X_{\mu}^k~
\Big| \CF( \{ X_{\nu}^k \} ) \Big|.
\label{absolute} 
\end{equation} 
This is algorithmically far more efficient, because now the problem
may again be reduced to the computation of the square of a Pfaffian,
which is readily available, avoiding the calculation of the Pfaffian 
itself.

The final result for the asymptotic eigenvalue densities as $\lambda
\rightarrow \infty$   supersymmetric systems in
$D=4,6,10$, for the supersymmetric system with gauge groups SU$(N)$, is 
\begin{equation} 
\mathbf{
\rho^{\rm SUSY}_D(\lambda) \sim \left\{ \begin{array}{cc} \lambda^{-3}
& \qquad D=4  \\ \lambda^{-7} & \qquad D=6  \\ \lambda^{-15} &
\qquad D=10 \end{array} \right.  }.
\label{power} 
\end{equation}
These laws were obtained by applying the above Markov chain random
walk tool to the measure eq.(\ref{absolute}), i.e.~we established
divergence of eq.(\ref{corr}) iff $k \geq 2,6,14$ (respectively
for $D=4,6,10$), leading to eq.(\ref{power}).  Note that these
power laws are independent of $N$. They demonstrate that the present
matrix models are very different from the classic Wigner-type
models. It would be interesting to obtain the generalization of
eq.(\ref{power}) to other gauge groups.

\acknowledgments
This work
was supported in part by the EU under Contract FMRX-CT96-0012.

\end{document}